\documentclass[a4paper,twocolumn,11pt,accepted=2023-06-30]{quantumarticle}
\pdfoutput=1

\usepackage{latexsym}
\usepackage{amsmath,amssymb,amsthm,thmtools}
\usepackage{mathtools,mathrsfs}
\usepackage{amsbsy}
\usepackage{soul}
\usepackage[caption=false]{subfig}
\usepackage{bbold}
\usepackage[numbers,sort&compress]{natbib}

\usepackage[utf8]{inputenc}

\usepackage[pdftex]{graphicx}
\graphicspath{{./figures/}}

\usepackage[usenames,dvipsnames,table]{xcolor}
\usepackage{float}

\usepackage[colorlinks=true]{hyperref}
\usepackage[nameinlink, capitalise]{cleveref}
\hypersetup{
  colorlinks   = true, 
  urlcolor     = green!80!black, 
  linkcolor    = blue, 
  citecolor    = red!80!black 
}

\usepackage{graphicx}

\usepackage[normalem]{ulem}

\usepackage{microtype}

\usepackage{physics}
\usepackage{easyReview}

\newcommand{\on}[1]{{\operatorname{#1}}}

\newcommand{\calE}{\mathcal{E}}
\newcommand{\calF}{\mathcal{F}}

\newcommand{\calS}{\mathcal{S}}

\newcommand{\parTitle}[1]{\noindent{\color{Mahogany}(\emph{#1})}}
 \renewcommand{\parTitle}[1]{}

\begin{document}

\title{
The meaning of redundancy and consensus in quantum objectivity
}

\author{Diana A.~Chisholm}
\affiliation{Universit\`a degli Studi di Palermo, Dipartimento di Fisica e Chimica -- Emilio Segr\`e, via Archirafi 36, I-90123 Palermo, Italy}
\affiliation{Centre for Quantum Materials and Technologies, School of Mathematics and Physics, Queen's University Belfast, BT7 1NN, United Kingdom}
\orcid{0000-0003-0496-888X}
\author{Luca Innocenti}
\affiliation{Universit\`a degli Studi di Palermo, Dipartimento di Fisica e Chimica -- Emilio Segr\`e, via Archirafi 36, I-90123 Palermo, Italy}
\orcid{0000-0002-7678-1128}
\author{G.~Massimo Palma}
\affiliation{Universit\`a degli Studi di Palermo, Dipartimento di Fisica e Chimica -- Emilio Segr\`e, via Archirafi 36, I-90123 Palermo, Italy}
\affiliation{NEST, Istituto Nanoscienze-CNR, Piazza S.~Silvestro 12, 56127 Pisa, Italy}
\orcid{0000-0001-7009-4573}
\begin{abstract}
    While the terms ``redundancy'' and ``consensus'' are often used as synonyms in the context of quantum objectivity, we show here that these should be understood as two related but distinct notions, that quantify different features of the quantum-to-classical transition.
    We show that the two main frameworks used to measure quantum objectivity, namely spectrum broadcast structure and quantum Darwinism,
    are best suited to quantify redundancy and consensus, respectively.
    Furthermore, by analyzing explicit examples of states with nonlocally encoded information, we highlight the potentially stark difference between the degrees of redundancy and consensus.
    In particular, this causes a break in the hierarchical relations between spectrum broadcast structure and quantum Darwinism.
    Our framework provides a new perspective to interpret known and future results in the context of quantum objectivity, paving the way for a deeper understanding of the emergence of classicality from the quantum realm.
\end{abstract}

\maketitle

\section{Introduction}

\emph{Introduction ---}
The emergence of classicality is arguably one of the oldest foundational open questions in quantum mechanics, and the quantum-to-classical transition still the subject of active research.
Decoherence~\cite{breuer_theory_2007,rivas_open_2011} is often considered a promising mechanism to explain such transition~\cite{schlosshauer_decoherence_2005,zurek_decoherence_2003}: as quantum systems are never really isolated~\cite{breuer_theory_2007, rivas_open_2011}, their surrounding environment ``reads'' the system, degrading coherent superpositions into statistical mixtures, and inducing the corresponding loss of quantum features~\cite{zurek_decoherence_2003}.
However, the theory of decoherence is not \textit{per se} sufficient to directly explain some pivotal markers of classicality, such as the emergence of \textit{objectivity} in quantum systems.

While an intrinsic property of classical systems, ``objectivity'' is usually understood in the quantum domain as consensus between observers.
More specifically, given a quantum system $\mathcal S$ interacting with an environment $\mathcal E$, we say that the system is objective when observers performing independent measurements on different 
subsets of the environment
reach a consensus about some property of the system~\cite{ollivier_objective_2004}. This is only possible if, due to prior interactions between system and environment, the relevant information was encoded with high \emph{redundancy} into the environment~\cite{ciampini_experimental_2018, chen_emergence_2019, unden_revealing_2019, chisholm_witnessing_2021}. 
It is usually accepted that the individual environmental constituents may not hold enough information about the system, so that it is necessary to group them together into environmental fractions $\calE_i$, and that each observer will measure a fraction instead of an environmental constituent.
In the literature, redundancy is used as a technical term to refer to how many times the information about the system is encoded into the environment, which is also intended to be the number of observers that can simultaneously access said information.
The notion of \emph{consensus} is also often found in the literature as a non-technical term, that usually refers to the presence of redundancy.

Redundancy and consensus have, however, clear English language meaning. In the context of quantum objectivity, those meanings suggest an interpretation of redundancy in terms of how many times the information about the system is encoded into the environment, and consensus as the number of observers that can simultaneously access said information.
Building from the intuitive notions of redundancy and consensus, we will provide rigorous operative definitions that reflect their meaning in the English language.

This will allow us to show that in several situations these concepts ought to be carefully distinguished,
and that redundancy is a necessary, but not sufficient, condition to achieve consensus.
In particular, we show that in the presence of nonlocal information encoding, consensus and redundancy quantify significantly different features.
In fact, even though some information about $\mathcal S$ might be redundantly encoded into the environment, it is possible that due to the nonlocal retrievability of such information, the observers might not be able to access it, and thus the redundancy might fail to realize into a corresponding amount of consensus.
Such situations arise naturally whenever there are inter-environmental interactions~\cite{ryan_quantum_2021,mirkin_many-body_2021, milazzo_role_2019, le_thermality_2021, giorgi_quantum_2015}.
While redundancy is an intrinsic property of a system-environment state, consensus is also a function of the particular scenario, and in particular of how the environment is distributed between observers.

Furthermore, we find that, remarkably, this approach leads naturally to the two main quantifiers of ``quantum objectivity'', namely, spectrum broadcast structures (SBS)~\cite{horodecki_quantum_2015} and quantum Darwinism (QD)~\cite{blume-kohout_quantum_2006, zurek_quantum_2009}. More precisely, we find QD and SBS to quantify consensus and redundancy, respectively.

Finally, we focus on the established hierarchy between SBS and QD: while SBS implies QD~\cite{horodecki_quantum_2015, korbicz_roads_2021}, the opposite is only true under the additional assumption of \textit{vanishing discord} and {\it strong independence}~\cite{le_strong_2019, korbicz_roads_2021, Feller_Comment_2021}.
We show that this hierarchy does not hold in scenarios where consensus and redundancy are distinct, a notable example being states with highly nonlocal information encoding.
In particular, SBS states need not satisfy the QD condition.

\begin{figure}
    \centering
    \includegraphics[width=\columnwidth]{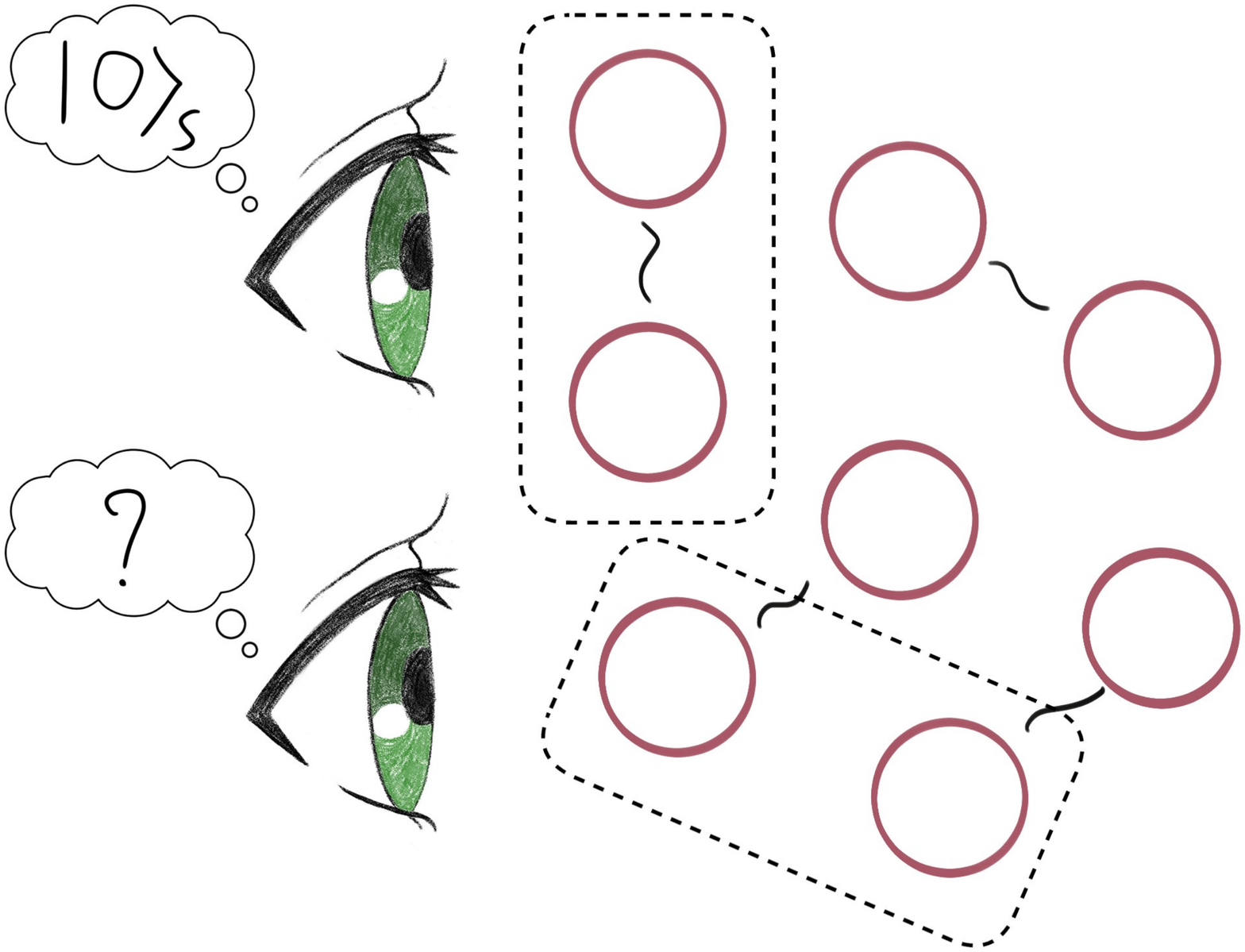}
    \caption{\textbf{Nonlocal information encoding.} When the system-environment state is the one in \cref{eq:generic_state}, then the information about the system is encoded into the correlations between the environmental constituents, and it can only be recovered by measuring all the qubits from the same group. This is an example where it is important to distinguish between the concepts of redundancy and consensus.
    }
    \label{fig:nonlocality}
\end{figure}

\section{Main results}
\textit{Redundancy and consensus ---}
To clarify the differences between the different aspects involved in the notion of \textit{quantum objectivity}, we introduce here operative definitions of ``redundancy'' and ``consensus'' in quantum systems.
Even though directly verifying these conditions might be hard in practice, we will show that they naturally reduce to QD and SBS in specific appropriate approximation regimes.

\textit{Definition of redundancy ---}
The idea of \textit{redundancy} is that information about the system is encoded into several independent environmental fractions.
More precisely, we thus say that a state $\rho$ has redundancy $n$, and write $\text{Red}(\rho)=n$, if $n$ is the largest integer such that there is a partition $\mathcal E=\bigotimes_{i=1}^n \mathcal E_i$ such that $I_{\rm acc}(\mathcal S:\mathcal E_i)\simeq S(\rho_\calS)$ for all $i=1,...,n$, where $S(\rho_\calS)$ is the von Neumann entropy of $\rho_\calS$, and $I_{\rm acc}(\calS :\calE_i)$ is the \textit{accessible} mutual information~\cite{preskill2016quantum} between system and $i$-th environmental fraction $\calE_i$.
The accessible information is here defined as the mutual information between the probability distributions resulting from measuring $\calS$ and $\calE_i$, maximized over all possible choices of measurements.
We use the accessible, rather than the full quantum mutual information, to ensure that correlations are actually observable from outcome probabilities.
Indeed, the numerical difference between the accessible information and the quantum mutual information takes the name of symmetric discord~\cite{ollivier_quantum_2001, henderson_classical_2001, rulli_global_2011}, and represent correlations that cannot be accessed through local measurements on the system and the environment. Therefore, witnessing a sufficient amount of quantum mutual information between the system and the environment may result in a false positive of objectivity as observers may not be able to access it~\cite{le_strong_2019}. A useful property of the accessible information is that it is bounded by the entropy of the system, that is $I_{\rm acc}(\calS:\calE_f)\leq S(\rho_\calS)$.

\textit{What redundancy does not quantify ---}
This notion of redundancy is however unconcerned with the practical retrievability of the information.
Even though information about a system can be encoded into its environment with high redundancy, actually accessing this information might require a careful partitioning of the environment, to avoid information being hidden in the correlations between different observers. 
To quantify the likelihood of actually observing information redundantly encoded into the environment, one ought to introduce the notion of \textit{consensus}.
While the redundancy is concerned with the maximum number of environmental fractions encoding information about $\calS$, the notion of consensus is also concerned with how hard it is to find partitions realizing a set amount of redundancy.
More precisely, we want to define the consensus of a state $\rho$ as the largest number of observers that can, \textit{with high probability}, retrieve information about the system from their measurement outcomes.

\textit{Definition of consensus ---}
To this end, consider the probability $P(\rho,n)$ of finding a partition $\calE=\bigotimes_{i=1}^n \calE_i$ with redundancy $n$:
\begin{equation}\label{eq:probability_for_consensus}\small
    P(\rho,n) \equiv
    \on{Prob}\big(
    I_{\rm acc}(\calS:\calE_i)\simeq S(\rho_\calS), \,\, i=1,...,n
    \big),
\end{equation}
where the probability effectively counts the number of partitions $\calE=\bigotimes_{i=1}^n \calE_i$ such that $\dim(\calE_i)=\dim(\calE_j)$ for all $i,j$ --- and thus such that $\dim(\calE_i)=f \dim(\calE)$ with $f=1/n$.
While in principle one could consider partitions with uneven fractions, we will focus on this simplest case, as it is the one usually considered in the topical literature.
We can now define the ``consensus'' of $\rho$ as 
\begin{center}
    the largest $n$ such that $P(\rho, n)\simeq1$.
\end{center}
Thus $\rho$ has a degree of consensus $n$ \textit{iff}, for any partition of the form $\calE=\bigotimes_{i=1}^n\calE_i$ with $\dim(\calE_i)=1/f$, each $\calE_i$ is maximally correlated with $\calS$, and the correlation is fully accessible~\footnote{It is possible to make the definitions of consensus and redundancy more mathematically rigorous by interpreting the equations $I_{\rm acc}(\calS:\calE_i)\simeq S(\rho_\calS)$ and $P(\rho,n)\simeq 1$ as standing for $I_\mathrm{acc}(\calS:\calE_i)\geq\calS(\rho_\calS)-\epsilon$ and $P(\rho, n)\geq1-\delta$, respectively, for suitable choices of thresholds $\epsilon$ and $\delta$.}.
From the above definitions follows that $P(\rho, n)\simeq1$ \emph{only if} $\operatorname{Red}(\rho)\geq n$. In other words, the consensus is never greater than the redundancy.

The quantity $P(\rho, n)$ is the probability that an observer is able, measuring a fraction of the environment of size $1/n$, to infer sufficient information about the system. 
In computing it, we assume that all the possible environmental fractions are sampled by the observers with equal probability. This is the simplest assumption one can make without going into the details of specific physical models. In specific scenarios, certain environmental fractions may be sampled with a higher probability than others (for example fractions with higher spatial connectivity), thus resulting in a different $P(\rho, n)$ than the one we obtain assuming equiprobable fractions.
However, even in this case our definition of consensus remains unchanged, as it does not depend on how $P(\rho, n)$ is computed.

In conclusion, redundancy is the maximum  number of observers that could, in principle, infer information about the system. Consensus is instead the guaranteed number of observers that can infer information about the system given a random partition of the environment. Assuming a one-shot random partition of the environment, we can be confident that each observer will be able to infer information about the system only if the number of fractions equals the value of consensus. However, by dividing the environment into a larger number of fractions, it is still possible for the fractions to be correlated with the system,  with the upper limit given by the redundancy. The number of observers being able to retrieve information about the system would then likely be higher than consensus, but lower than redundancy.
Therefore, while we argue that objectivity should be measured using consensus, both redundancy and consensus capture two different, nonequivalent aspects of the objectivity of a system.

\section{Relations with known measures of objectivity}

\textit{Non-equivalence of redundancy and consensus ---}
This formalization of the notions of redundancy and consensus makes it clear that they quantify distinct features of the emergence of quantum objectivity.
In particular, it is possible to have high redundancy and low consensus, in situations where the redundant information encoding cannot be easily accessed by the observers without taking great care in how the environmental fractions are distributed.
Furthermore, we will show here that the two main quantifiers of quantum objectivity used in the literature, SBS and QD, directly correspond to redundancy and consensus, respectively.

\textit{Redundancy vs SBS ---}
An SBS state is one that admits a decomposition of the form
\begin{equation}
    \rho_{\rm SBS} =
    \sum_i p_i \ketbra{i}{i}\otimes\bigotimes_{j=1}^N R_i^j,
\end{equation}
for some collection of states $R_i^j$ such that $R_{i}^{j}R_{i'}^{j}=\delta_{ii'}(R_{i}^{j})^{2}$ for all $j$~\cite{horodecki_quantum_2015}.
Such a decomposition is defined with respect to a specific partition $\calE=\bigotimes_{j=1}^N\calE_j$, of the environment, $R_i^j\in \calE_j$, and the conditions on $R_i^j$ ensure that different states of the system can be recovered from measurements in each fraction $\calE_j$.
SBS states are thus always objective, provided observers are able to measure environmental fractions compatibly with the partitioning corresponding to the SBS structure.
It follows that the redundancy of such a $\rho_{\rm SBS}$ is at least $N$. Furthermore, if N is the largest number of environmental fractions that allow to write the state as an SBS, then the redundancy of such state would be precisely N.

\textit{Consensus vs QD ---}
QD defines ``objectivity'' via the quantum mutual information (QMI) between system $\calS$ and an environment fraction $\calE_f$ of size $f\dim(\calE)$, where $f\in(0,1)$~\cite{blume-kohout_quantum_2006, zurek_quantum_2009}.
In many relevant scenarios~\cite{mironowicz_monitoring_2017,lorenzo_anti-zeno-based_2020,megier_correlations_2022, touil_eavesdropping_2022, roszak_entanglement_2019, roszak_glimpse_2020, cakmak_quantum_2021}, the systems encodes information uniformly in each of the environmental constituents, therefore, all environmental fractions of the same size will hold the same amount of information about the system.
For non-uniform environments~\cite{chisholm_witnessing_2021, blume-kohout_simple_2005, le_objectivity_2018, ryan_quantum_2021, milazzo_role_2019, mirkin_many-body_2021, Zwolak_Redundancy_2017}, the QMI will instead depend on the choice of environmental fraction. In such instances, it is therefore warranted to consider the \textit{average} QMI, defined as
$\tilde{I}(\calS:\calE_f) \equiv
    \left\langle
    I(\calS:\calE_f)
    \right\rangle$,
where the average is taken with respect to all environmental fractions of the same size $\calE_f$~\cite{blume-kohout_simple_2005, le_objectivity_2018, chisholm_witnessing_2021, Zwolak_Redundancy_2017}.
More explicitly, this average QMI can be written as
\begin{equation}
    \tilde I(\calS:\calE_f)\equiv \frac{1}{\mathcal{N}}
    \sum_{\calF}
    I(\calS:\calF),
    \label{eq:average_QMI}
\end{equation}
where the sum is over all possible $\mathcal{N}=\binom{\dim(\calE)}{f\dim(\calE)}$ subspaces $\calF\le \calE$ with $\dim(\calF)=f\dim(\calE)$.
\Cref{eq:average_QMI} shows that the averaged QMI is computed by assuming equiprobability for all the possible environmental fractions.
This is standard practice in the literature~\cite{blume-kohout_simple_2005, Zwolak_Redundancy_2017, le_objectivity_2018}, although the definition can in principle be extended by assigning a more general probability distribution to the different environmental fractions.

A state is then said to be objective according to QD if there is $f\in(0,1)$ such that
\begin{equation}\label{eq:Darwin}
    \tilde I(\calS:\calE_f)\simeq S(\rho_\calS).
\end{equation}
If $f$ satisfies~\cref{eq:Darwin}, then there are at least $1/f$ observers that can simultaneously agree on some property of the system.
We can then quantify \textit{QD-objectivity} as $1/f$ for the smallest such $f$.
One further possible issue arising in the context of QD is the presence of \textit{quantum discord}~\cite{ollivier_quantum_2001, wiseman_quantum_2013} in these QMI. Namely, in some scenarios computing $I(\calS:\calE_f)$ might falsely overestimate the actual accessible correlations between system and environment, which are the correlations that can be observed via some suitable choice of measurement basis on the environmental fractions. These arguably are, for the purpose of objectivity, the correlations one is actually interested in, and therefore the use of \textit{accessible} mutual informations might more accurately quantify the sought-after objectivity of states, this is sometimes referred to as strong quantum Darwinism~\cite{le_strong_2019, le_witnessing_2020}.
For these reasons, we will refer from now on to QD-objectivity as computed in terms of the accessible information.

We can now observe that QD-objectivity precisely corresponds to the degree of consensus previously introduced.
In fact, if $P(\rho,n)\simeq 1$, then any partitioning of the environment of the form $\calE=\bigotimes_{i=1}^n\calE_i$ gives $I_{\rm acc}(\calS:\calE_i)\simeq S(\rho_\calS)$, and therefore the average accessible mutual information also satisfies $\tilde I_{\rm acc}(\calS:\calE_f)\simeq S(\rho_\calS)$ with $1/f=n$.
Vice versa, if $\tilde I_{\rm acc}(\calS:\calE_f)\simeq S(\rho_\calS)$, then each $\calE_i$ gives $I_{\rm acc}(\calS:\calE_i)\simeq S(\rho_\calS)$,
and thus $P(\rho,n)\simeq1$. This is because $\max\left(I_{\rm acc}(\calS:\calE_f)\right) \leq S(\rho_\calS)$.

It is worth noting that this equivalence between QD-objectivity and consensus hinges on the definition of $P(\rho,n)$ as the probability that a random partition with equal sized fractions gives maximal correlations.
If one were instead to consider, for example, the probability of a \textit{completely random} partition having $I_{\rm acc}(\calS:\calE_i)\simeq S(\rho_\calS)$, then $P(\rho,n)\simeq 1$ would imply that even partitions involving fractions with single qubits would necessarily have to be maximally correlated to the system.
While such situations are in principle possible, the associated quantifier would differ substantially from what QD-objectivity measures, which is why we stick to the cases with equal-sized fractions here.

\section{States highlighting the difference between consensus and redundancy}

\emph{Nonlocal information encoding ---}
A notable class of states highlighting the differences between redundancy and consensus are those in which the information about the system can only be recovered by means of collective measurements on several environmental constituents, as pictorially represented in~\mbox{\cref{fig:nonlocality}}.
While it is usually the case that a single environmental constituent does not hold enough information about the system, we refer here to the fragility of the information encodings with respect to loss of pieces of the environment, and show that it is tightly related with the departures between redundancy and consensus.
Here, fragility is not to be intended as a precisely definite concept, but rather as the fact that the amount of information an environmental fraction holds about the system quickly degrades when even a small part of said fraction is lost.
We focus in particular on the extreme cases, where losing even a single environmental constituent results in a severe degradation of the correlations.

We focus here on many-qubit states for simplicity.
To have both high redundancy and maximally fragile encoded information, we require the state of the system to be maximally correlated with a number of environmental states of the form
\begin{equation}\label{eq:GHZ_states}
    |\on{GHZ}^{(k)}_\pm\rangle
    \equiv
    \frac1{\sqrt 2}(
        \ket0^{\otimes k} \pm \ket1^{\otimes k}
    ) \in (\mathbb{C}^2)^{\otimes k}.
\end{equation}
While these two states are fully distinguishable when all $k$ qubits are measured, they become completely indistinguishable if even only a single qubit is lost, as
\begin{equation}
    \tr_i( \on{GHZ}_+^{(k)}) =
    \tr_i( \on{GHZ}_-^{(k)}), \,\,\forall i=1,...,k.
\end{equation}
Note that GHZ-like states are characterized by not being determined by their reduced density matrices. More explicitly, this means that for any pair of orthogonal states $\ket\psi,\ket\phi\in(\mathbb{C}^2)^{\otimes k}$ such that $\tr_i(\ketbra\psi)=\tr_i(\ketbra\phi)$ for all $i$, $\ket\psi$ and $\ket\phi$ are local unitary (LU) equivalent to $\ket*{\on{GHZ}_\pm^{(k)}}$~\cite{walck_only_2008, walck_only_2009}. GHZ-like states are therefore the ones that result in the highest degree of fragility when used to encode information.

\begin{figure}
    \centering
    \includegraphics[trim={2cm 0cm 2cm 0cm}, clip, width=\columnwidth]{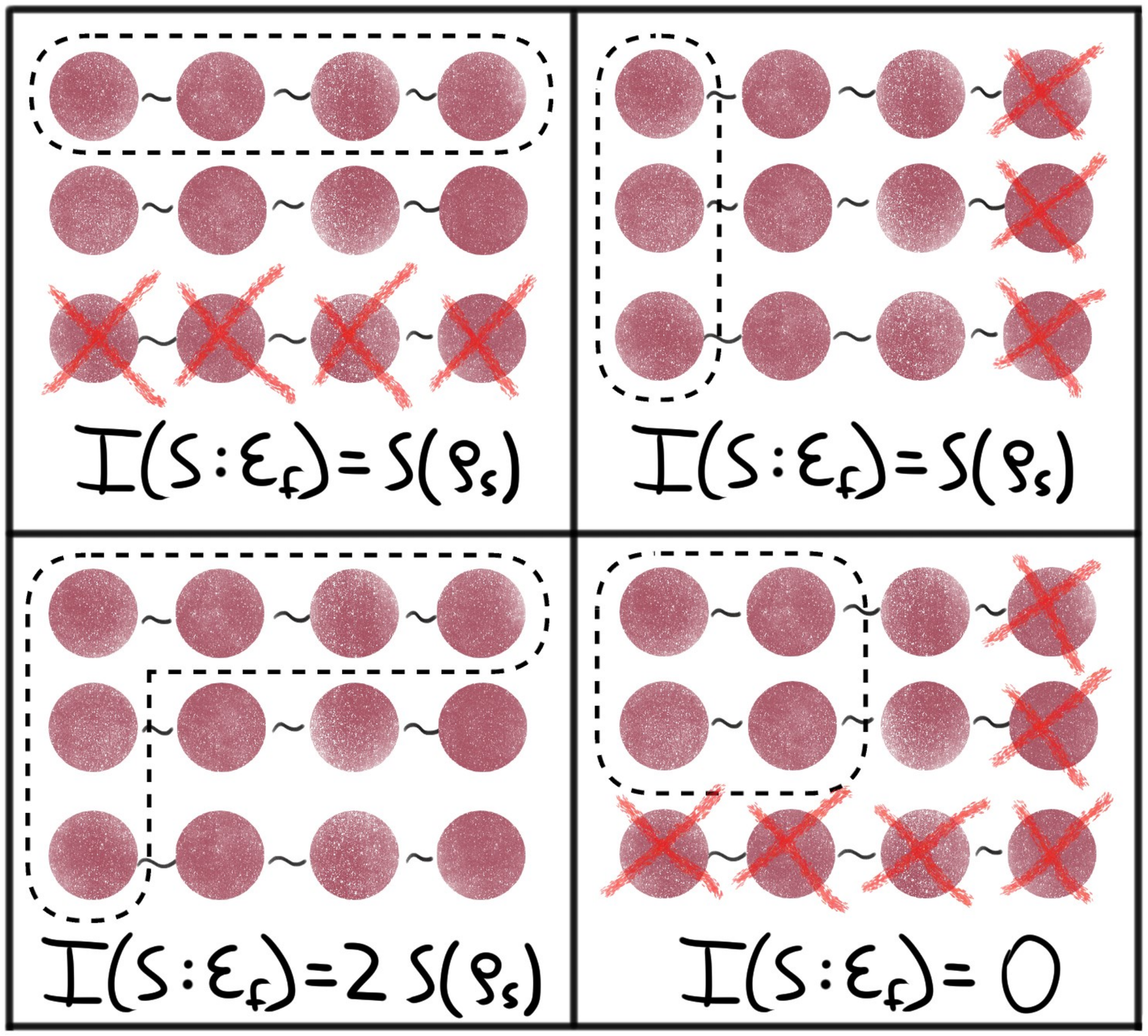}
    \caption{
    \textbf{How measuring different qubits affects the QMI}.
    In each figure, the three rows correspond to the three $\ket*{\on{GHZ}_\pm^{(k)}}$ macrofractions in the state $\ket{\Psi}_{k,N}$, with degree of nonlocality $k=4$ and redundancy $N=3$.
    The disks corresponds to the physical environmental qubits.
    In each scenario, depending on which subset of qubits is being measured, we get a different corresponding QMI.
    The dashed lines highlight qubits that are being measured, and the red crosses mark ones that have been traced out.
    Measuring qubits which are neither encircled nor crossed out does not affect the QMI.
    Measuring a full row (\textbf{\textit{top left}}) or a full column (\textit{\textbf{top right}}) we get $I(S:\calE_f)=S(\rho_\calS)$, while measuring both a full row and a full column (\textit{\textbf{bottom left}}), $I(S:\calE_f)=2S(\rho_\calS)$. If neither are true (\textbf{\textit{bottom right}}), there is no correlation between system and the measured environmental fraction, corresponding to $I(S:\calE_f)=0$.
    }
    \label{fig:measure}
\end{figure}

\begin{figure}[t]
    \centering
    \includegraphics[width=\columnwidth]{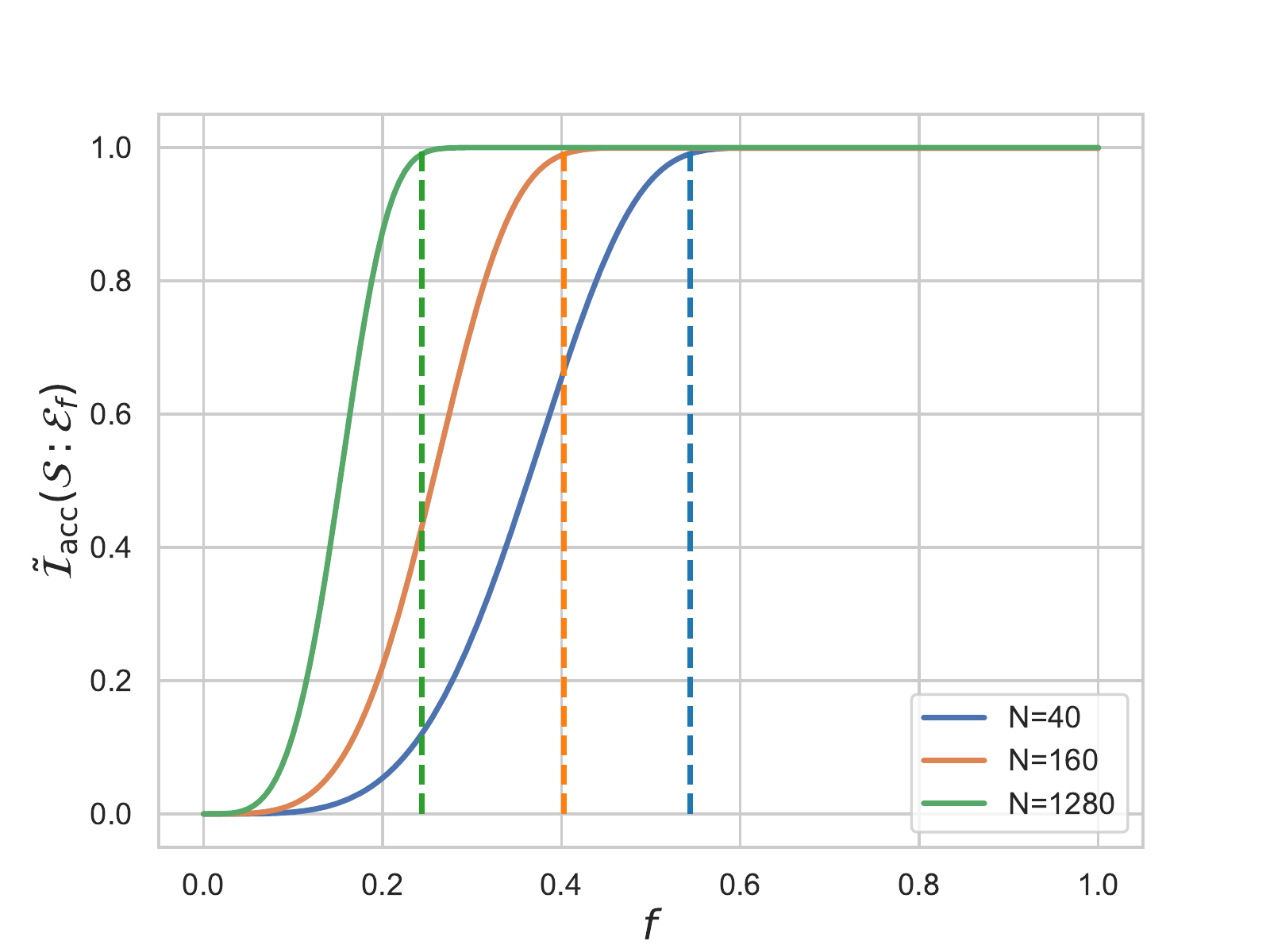}
    \caption{\textbf{Average accessible mutual information between system and environmental fraction as a function of the fraction size $f$}. The system-environment state is the one depicted in \cref{eq:generic_state}, the three different curves correspond to different values of redundancy $N$. The vertical bars correspond to the minimum value $f_0$ such as $I(\calS:\calE_{f_0})=0.99S(\rho_\calS)$, which are needed to compute consensus as $1/f_0$.
     Notice how high redundancy does not always result in the typical objectivity plateau of quantum Darwinism. In particular, redundancy values of 40, 160 and 1280 correspond to consensus values of 1, 2 and 4, respectively. For all cases $k=4$.}
    \label{fig:increase}
\end{figure}

We thus consider system-environmental states $\ket{\Psi}_{k,N}\in(\mathbb{C}^2)^{Nk+1}$ of the form
\begin{equation}\label{eq:generic_state}\small
    \ket\Psi_{k,N} \equiv \frac{1}{\sqrt2}(
    \ket0\otimes \ket*{\on{GHZ}_+^{(k)}}^{\otimes N} +
    \ket1\otimes \ket*{\on{GHZ}_-^{(k)}}^{\otimes N}
    ).
\end{equation}
These states have degree of redundancy $N$.
If furthermore $k=1$, the information is encoded locally, and we also have a degree of consensus $N$.
In fact, in this case, there is only one possible partitioning of the environment --- which is composed of $Nk=N$ qubits --- into $N$ fractions, so that $P(\rho,N)=1$.

However, as soon as $k>1$, the two aspects of objectivity diverge.
In such instances, the QMI depends on the way in which the $kN$ environmental qubits are partitioned between observers.
We recognize in particular two conditions that are sufficient to determine the QMI $I(\calS:\calE_i)$ corresponding to a given environmental fraction $\calE_i$.
The first condition is whether at least one full GHZ state is fully contained in $\calE_i$.
The second condition is whether at least one qubit from each GHZ state is in $\calE_i$.
If neither condition is satisfied $I(\calS:\calE_i)=0$; if either condition is satisfied $I(\calS:\calE_i)=S(\rho_\calS)$; finally, if both conditions are satisfied, then $I(\calS:\calE_i)=2S(\rho_\calS)$.
The corresponding four possible scenarios are pictorially represented in~\cref{fig:measure}.
\begin{figure*}[t]
    \centering
    {\includegraphics[trim={0cm 0.5cm 0cm 0.2cm}, clip,width=\textwidth]{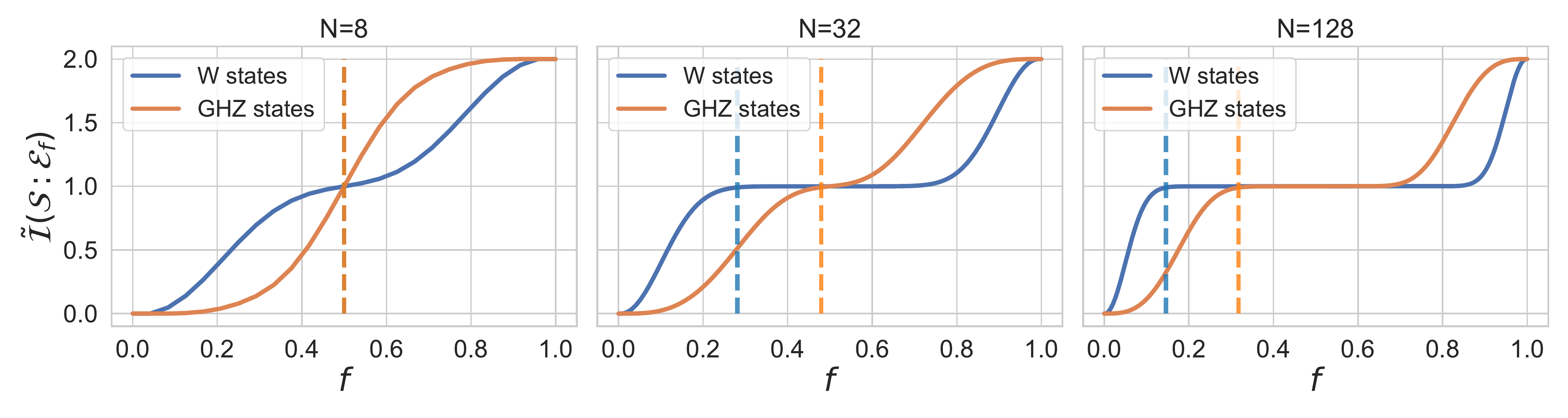}}
    \caption{
    \textbf{Averaged QMI encoding the information in GHZ states and W states.}
    While the overall effects of nonlocal information encoding remain unaltered, encoding information in W states always results in higher consensus compared to GHZ states.
    In each figure $k=3$, while the redundancy is $N=8, N=32$, and $N=128$, respectively.
    GHZ states result in consensus values of 1, 2 and 3 respectively, while W states result in consensus values of 1, 3 and 6 respectively.
    }
    \label{fig:H_W}
\end{figure*}
We refer to the supplementary calculations (SC) for the full derivation of these quantities, where we also show that performing the same calculations for the accessible mutual information gives $I_{\rm acc}(S:\calE_i)=S(\rho_\calS)$ in all cases except when neither condition is satisfied.
With these results, computing average QMI and accessible mutual information reduces to a combinatorial problem.

In~\cref{fig:increase} we give the values of $\tilde I_{\rm acc}(\calS:\calE_f)$ as a function of $f$ for different values of $N$, for $k=4$.
As clear from these results, a large degree of redundancy does not guarantee an equal amount of consensus. Therefore, it could be possible that, on average, observers will not measure any intact GHZ state, and will be unable to recover information about the system, rendering the redundant information inaccessible. 
While the value of redudnancy simply coincides with the number $N$, consensus is evaluated by choosing a threshold value for the accessible information and then finding the minimum value $f_0$ such that $I_{\rm acc}(\calS:\calE_{f_0})$ is equal or greater than the chosen threshold. Consensus is then given by $1/f_0$ approximated to the smaller integer number. Throughout the paper we choose our threshold value to be $0.99S(\rho_\calS)$.
In particular, redundancy values of 40, 160 and 1280 correspond to consensus values of 1, 2 and 4.
This choice of states thus highlights how nonlocal information encoding results in non-equivalence between redundancy and consensus.
We also see that the consensus increases with the size of the environment, which is a consequence of the fact that for large $N$ the probability of selecting at least $k\ll N$ qubits belonging to the same GHZ state tends to unity.
In the macroscopic limit, the system exhibits objectivity even in the worst-case scenario of maximally nonlocal information encoding.
As discussed previously, it is possible to recover information about the system by measuring a single element from each $\ket{\on{GHZ}}$ --- that is, reading each element of a column in~\cref{fig:measure}. However, in the macroscopic limit of $N\gg k$, the probability of picking such partition goes to zero.

\emph{Different information encoding: W states.—}
We have shown how states that are LU equivalent to the GHZ state highlight a stark departure between redundancy and consensus.
Nonetheless, many entangled states also encode information with a high degree of fragility.
One such standard example of three-qubit states are W states~\cite{Dur_Three_2000}.
These can be seen as an intermediate case between the GHZ case and the case in which information about the system is encoded into each individual environmental qubit.
We then consider system-environment states of the form
\begin{equation}\label{eq:W_encoding}
    \frac12(
    \ket*0\otimes\ket*{W_+^{(k)}}^{\otimes N} +
    \ket*1\otimes\ket*{W_-^{(k)}}^{\otimes N}
    ),
\end{equation}
with $\ket*{W_\pm^{(k)}}$ a pair of $W$-like orthogonal many-qubit states.
In the case of $k=3$, these read
\begin{equation}\label{eq:W_states}
\begin{gathered}
    \ket*{W_+^{(3)}} \equiv \frac{1}{\sqrt3}(
    \ket{001} + \ket{010} + \ket{100}
    ), \\
    \ket*{W_-^{(3)}} \equiv \frac{1}{\sqrt3}(
    \ket{001} + \omega_3 \ket{010} + \omega_3^2\ket{100}
    ),
\end{gathered}
\end{equation}
with $\omega_3\equiv e^{2\pi i/3}$.
These definitions are straightforwardly generalized to general $k$.

In this case, the mutual information between system and environmental fraction does not reduce to the four possible cases as with the GHZ state structure, as now it becomes relevant how many qubits are selected from each W state. In any case, it is still possible to reduce the QMI calculation to a combinatorial, albeit more complex, problem. See the SC for the full calculation.

While the individual qubits still carry no information about the system, a partially surviving $\ket*{W_+^{(k)}}$ state is slightly distinguishable from the $\ket*{W_-^{(k)}}$ counterpart, meaning that an observer may be able to determine the state of the system without having access to a full W state. For this reason we expect it to be easier to achieve a higher degree of consensus in this case compared to the GHZ case.

This is confirmed by~\cref{fig:H_W}, where we compare the average QMI between W-state and GHZ-state encodings for a fragility degree of $k=3$. We can see that the W-state encoding always results in a higher QMI compared with the GHZ case. 
In particular, redundancy values of 8, 32 and 128 correspond to consensus values of 1, 3 and 6 for W-states and 1, 2 and 3 for GHZ-states.
In any case, even though a W-state encoding results in a higher degree of consensus given an equal amount of redundancy, the qualitative behaviour and relative implications are the same as with the GHZ case.

\section{Conclusions}

We have shown that objectivity in quantum systems can be quantified as redundancy or consensus, two different notions of which we have given operative definitions.
We further showed that these two notions can be quantified by means of SBS and QD, respectively.
While redundancy and consensus often reduce to the same quantity, they can be very different, in particular in scenarios where the information about the system is encoded in a highly nonlocal way, such as when the information is encoded in GHZ or W states.
Our results highlight the role of nonlocally encoded information in the emergence of objectivity, and show that in general it tends to hinder the emergence of consensus. This is due to the fact that, even if redundancy is high, it will be difficult for observers to extract relevant information about the system.
On the other hand, based on probabilistic considerations, the hindering effects of such nonlocal information encoding is compensated by large environments, which suggests that, in the limit of macroscopic environments, high values of consensus are reached even with nonlocal information encoding.

Finally, we discussed how the difference between redundancy and consensus breaks the previously established relations between SBS and QD, since there are wide classes of states where being SBS does not necessarily imply QD, although it should be noted that the use of the averaged, accessible information in the QD approach is a fundamental requirement for the definition of consensus.
States can have a degree of redundancy much higher than consensus, and therefore measuring a certain degree of redundancy via SBS does not imply a similar degree of QD-objectivity.
We note however, that said relation could be re-established, provided that some extra care is taken whenever redundancy and consensus do not coincide: if a state is SBS for (on average) all environmental partitionings of a certain minimum size, then SBS would imply QD even in the nonlocal scenario.
Our framework allows for a clearer interpretation of results obtained using quantifiers such as SBS and QD, thus paving the way for a more thorough understanding of the different aspects of quantum objectivity, and thus the emergence of classicality from the quantum world.
Future venues of research include an analysis of redundancy and consensus across diverse state classes, and the exploration of how our results might change if we were to introduce a biased distribution when dividing the environment among the different observers in the process of defining consensus.
Another interesting aspect that might warrant further investigation is a thorough quantitative analysis of the relation between fragility of the information encoding and departure between redundancy and consensus. While in this paper we focused exclusively on some notable cases such as GHZ and W states, it might be possible to leverage the general framework of quantum error correction to define quantitatively a suitable notion of fragility, and tie it to the quantum objectivity properties of the corresponding state.

\acknowledgments
D.A.C.~acknowledges support from the Horizon Europe EIC Pathfinder project  QuCoM (Grant Agreement No.101046973). L.I.~acknowledges support from MUR and AWS under project PON Ricerca e Innovazione 2014-2020, “calcolo quantistico in dispositivi quantistici rumorosi nel regime di scala intermedia” (NISQ - Noisy, IntermediateScale Quantum). G.M.P.~acknowledges support from MUR through project PRIN (Project No.2017SRN-BRK QUSHIP). .

\bibliographystyle{plainnat}
\bibliography{Ref}

\onecolumngrid
\clearpage
\appendix
\setcounter{equation}{0}
\renewcommand{\theequation}{\thesubsection S.\arabic{equation}}

\section*{{\huge Supplementary calculations}}

Here we show how to obtain the average mutual information between system $\calS$ and environmental fraction $\calE_f$, when the global system-environment state is the one in~\cref{eq:generic_state} (information encoded in GHZ-like states) and~\cref{eq:W_encoding} (information encoded in W-like states). In both cases, we will first have to compute all possible values of $I(\calS:\calE_f)$, which depend on the environmental constituents making up $\calE_f$. We will then make combinatorial considerations to compute the probability associated with each $I(\calS:\calE_f)$ value, thus computing the averaged mutual information. The obtained results correspond to the ones shown in~\cref{fig:increase} and~\cref{fig:H_W}.

\section{Mutual information for GHZ encodings}

    Here we consider all possible states resulting from tracing out environmental components from the global state, when information about the system is encoded in GHZ-like states, and compute the QMI between the system and the resulting environmental fraction. We are interested in computing the accessible information, but we will see how, in this specific case, we can easily recover the accessible information by computing the QMI first.
    The global state is the same as in~\cref{eq:generic_state}
    \begin{equation}
        \ket{\Psi}=\frac{1}{\sqrt{2}}\left(\ket{0}\otimes\ket{\tilde{+}}^{\otimes N}+\ket{1}\otimes\ket{\tilde{-}}^{\otimes N}\right),
    \end{equation}
    where $\ket{\tilde{\pm}}=\frac1{\sqrt 2}(\ket0^{\otimes k} \pm \ket1^{\otimes k})$ are the same as in~\cref{eq:GHZ_states}, the GHZ states into which the information is encoded.
    
    Depending on which qubits are traced out from the global state, four different scenarios may arise, which we will analyse individually.
    
    \subsection{First case}
    So long as we trace out entire GHZ states, so that only $fN$ remain ($f\in(0, 1)$), the resulting state is the following:
     \begin{equation}
        \rho_{\mathcal{S}\mathcal{E}_f^{(0)}}=\frac{1}{2}\left(\ket{0}\bra{0} \otimes \ket{\tilde{+}} \bra{\tilde{+}}^{\otimes fN} +\ket{1}\bra{1}\otimes \ket{\tilde{-}} \bra{\tilde{-}}^{\otimes fN}\right),
        \label{Seq:10}
    \end{equation}
    whose entropy is $1$ bit.
    If we trace out the system as well we obtain the following
    \begin{equation}
        \rho_{\mathcal{E}_f^{(0)}}=\frac{1}{2}\left(
        \ket{\tilde{+}}
        \bra{\tilde{+}}^{\otimes fN} + \ket{\tilde{-}}
        \bra{\tilde{-}}^{\otimes fN}\right),
    \end{equation}
    whose entropy is also $1$. Hence the mutual information is $I(\calS:\calE_f^{(0)})=1$.
    
    Let's consider the case in which we trace out only some qubits from a GHZ state. We can rewrite~\cref{Seq:10} as follows
    \begin{equation}
        \rho_{\mathcal{S}\mathcal{E}_f^{(0)}}=\frac{1}{2}\left(
        \ket{0}\bra{0} \otimes 
        \ket{\tilde{+}}\bra{\tilde{+}}^{\otimes fN-1}\otimes
        \ket{\tilde{+}}\bra{\tilde{+}} +\ket{1}\bra{1}\otimes 
        \ket{\tilde{-}} \bra{\tilde{-}}^{\otimes fN-1}\otimes\ket{\tilde{-}}\bra{\tilde{-}}\right),
    \end{equation}
    and after tracing out $m$ ancillae from the same group (with $m < k$) we obtain
    \begin{equation}
        \rho_{\mathcal{S}\mathcal{E}^{(0)}_f}=\frac{1}{4}\left(\ket{0}\bra{0} \otimes\ket{\tilde{+}} \bra{\tilde{+}}^{\otimes fN-1} +\ket{1}\bra{1}\otimes \ket{\tilde{-}} \bra{\tilde{-}}^{\otimes fN-1}\right)
        \otimes\left(\ket{0}\bra{0}^{\otimes k-m}+\ket{1}\bra{1}^{\otimes k-m}\right).
    \end{equation}
    This is the tensor product of two states of entropy $1$, meaning it itself has entropy $2$.
    Similarly after tracing out the system we have     
    \begin{equation}
        \rho_{\mathcal{E}_f^{(0)}}=\frac{1}{4}\left( \ket{\tilde{+}}\bra{\tilde{+}}^{\otimes fN-1} + 
        \ket{\tilde{-}}\bra{\tilde{-}}^{\otimes fN-1}\right)\otimes\left(\ket{0}\bra{0}^{\otimes k-m}+\ket{1}\bra{1}^{\otimes k-m}\right),
    \end{equation}
    also of entropy $2$.
    The procedure can be generalized: so long as we trace individual qubits from a complete GHZ state, we increase the entropy of $\mathcal{S}\mathcal{E}_f^{(0)}$ and $\mathcal{E}_f^{(0)}$ by one. The mutual information is $I(\calS:\calE_f^{(0)})=1$ as long as there is still an intact GHZ state remaining.
    \subsection{Second case}
    The previous results correspond to the case were at least one GHZ state is entirely traced out, and that at least one GHZ state remained intact.
    Let's now consider the case where we trace out a single qubit from the global (pure) state. The initial state is:
    \begin{equation}
        \rho_{\calS\calE}=\frac{1}{2}\Big[\ket{0}\otimes\ket{\tilde{+}}^{\otimes N-1}\otimes\ket{\tilde{+}}+\ket{1}\otimes\ket{\tilde{-}}^{\otimes N-1}\otimes\ket{\tilde{-}}\Big]\Big[\bra{0}\otimes\bra{\tilde{+}}^{\otimes N-1}\otimes\bra{\tilde{+}}+\bra{1}\otimes\bra{\tilde{-}}^{\otimes N-1}\otimes\bra{\tilde{-}}\Big],
    \end{equation}
    by tracing out a single environmental qubit we obtain
    \begin{equation}
    \begin{split}
        \rho_{\calS\calE_f^{(1)}}=&\frac{1}{2}\Big[\ket{0}\otimes\ket{\tilde{+}}^{\otimes N-1}+\ket{1}\otimes\ket{\tilde{-}}^{\otimes N-1}\Big]\Big[\bra{0}\otimes\bra{\tilde{+}}^{\otimes N-1}+\bra{1}\otimes\bra{\tilde{-}}^{\otimes N-1}\Big]\otimes\ket{0}\bra{0}^{\otimes k-1}\\
        +&\frac{1}{2}\Big[\ket{0}\otimes\ket{\tilde{+}}^{\otimes N-1}-\ket{1}\otimes\ket{\tilde{-}}^{\otimes N-1}\Big]\Big[\bra{0}\otimes\bra{\tilde{+}}^{\otimes N-1}-\bra{1}\otimes\bra{\tilde{-}}^{\otimes N-1}\Big]\otimes\ket{1}\bra{1}^{\otimes k-1},
    \end{split}
    \end{equation}
    this is a state of entropy $1$. If we trace out the system as well we obtain
    \begin{equation}
    \rho_{\calE_f^{(1)}}=\frac{1}{4}\Big[\ket{\tilde{+}}\bra{\tilde{+}}^{\otimes N-1}+\ket{\tilde{-}}\bra{\tilde{-}}^{\otimes N-1}\Big]\otimes\Big[\ket{0}\bra{0}^{\otimes k-1}+\ket{1}\bra{1}^{\otimes k-1}\Big],
    \end{equation}
    that is instead of entropy $2$, meaning that the mutual information is $I(\calS:\calE_f^{(1)})=2$ even if the global state is not pure.
    We write the generic state in the case where $n$ qubits, pertaining to different GHZ groups, have been traced out
    \begin{equation}\footnotesize
        \rho_{\calS\calE_f^{(1)}}=\frac{1}{2^n}\sum_{a_{1}=0}^{1}\dots\sum_{a_{n}=0}^{1}\Big[\ket{0}\otimes\ket{\tilde{+}}^{\otimes N-n}+(-)^{\sum a_{i}}\ket{1}\otimes\ket{\tilde{-}}^{\otimes N-n}\Big]\Big[\bra{0}\otimes\bra{\tilde{+}}^{\otimes N-n}+(-)^{\sum a_{i}}\bra{1}\otimes\bra{\tilde{-}}^{\otimes N-n}\Big]\bigotimes_{i=0}^{n}\ket{a_{i}}\bra{a_{i}}.
    \end{equation}
    Since all the terms in the sum are orthogonal with one another, we have a statistical mixture of $2^{n}$ states, all with equal probability, so that the entropy of this state is $n$.
    By tracing out the system we obtain
    \begin{equation}
        \rho_{\calE_f^{(1)}}=\frac{1}{2^{n+1}}\Big[\ket{\tilde{+}}\bra{\tilde{+}}^{\otimes N-n}+\ket{\tilde{-}}\bra{\tilde{-}}^{\otimes N-n}\Big]\sum_{a_{1}=0}^{1}\dots\sum_{a_{n}=0}^{1}\bigotimes_{i=0}^{n}\ket{a_{i}}\bra{a_{i}},
    \end{equation}
    that instead has entropy $n+1$.
    This means that even as we trace out environmental qubits, as long as me make sure that no GHZ state is traced out entirely, the mutual information between system and environment is $I(\calS:\calE_f^{(1)})=2$.
    \subsection{Third case}
    We can also consider the scenario when qubits are traced out from all GHZ states but no GHZ state is traced out entirely, in this case the resulting state is
    \begin{equation}
        \rho_{\calS\calE_f^{(2)}}=\frac{1}{2^N}\sum_{a_{1}=0}^{1}\dots\sum_{a_{n}=0}^{1}\Big[\ket{0}+(-)^{\sum a_{i}}\ket{1}\Big]\Big[\bra{0}+(-)^{\sum a_{i}}\bra{1}\Big]\bigotimes_{i=0}^{N}\ket{a_{i}}\bra{a_{i}},
    \end{equation}
    of entropy $N$, and when tracing out the system we obtain
    \begin{equation}
        \rho_{\calE_f^{(2)}}=\frac{1}{2^N}\sum_{a_{1}=0}^{1}\dots\sum_{a_{n}=0}^{1}\bigotimes_{i=0}^{N}\ket{a_{i}}\bra{a_{i}},
    \end{equation}
    Also of entropy $N$, meaning that in this case the mutual information is $I(\calS:\calE_f^{(2)})=1$.
    \subsection{Fourth case}
    There is only one scenario we have yet not considered, the case in which at least one qubit has been traced out from each GHZ state, and at least one GHZ state has been traced out entirely. in this case system and environment are in a tensor product, meaning that there are no correlations and that the mutual information is $I(\calS:\calE_f^{(3)})=0$:
    \begin{equation}
        \rho_{\mathcal{S}\mathcal{E}_f^{(3)}}=\frac{1}{2^{n+1}}\left( \ket{0}\bra{0} + 
        \ket{1}\bra{1}\right)\otimes\left(\ket{0}\bra{0}^{\otimes k-m}+\ket{1}\bra{1}^{\otimes k-m}\right)^{\otimes n}.
    \end{equation}

    For simplicity, we have considered the case where the global state is in an equal weight superposition between $\ket{0}\otimes\ket{\tilde{+}}^{\otimes N}$ and $\ket{1}\otimes\ket{\tilde{-}}^{\otimes N}$. All the calculations remain correct even for different superposition weights, as long as the resulting mutual information is not measured in bits, but in units of the system entropy.
    
    Here we have computed the quantum mutual information. If one is interested in the accessible mutual information, it is worth noting how, whenever the quantum mutual information between system and environment is 1, the corresponding states are classical-classical, meaning that the accessible information is also 1. This also implies that when the quantum mutual information is 2, the accessible information is 1.

\section{Combinatorial computation for GHZ encodings}
Following the considerations in the previous section, we can see that, given a specific environmental fraction $\calE_f$, there are two conditions that define the mutual information $I(\calS:\calE_f)$.
\begin{itemize}
    \item Condition A: $\calE_f$ contains at least one complete GHZ state
    \item Condition B: $\calE_f$ contains at least one qubit from each GHZ state
\end{itemize}
If only one of the two conditions is satisfied, then $I(\calS:\calE_f)=S(\rho_\calS)$. If both conditions are satisfied then $I(\calS:\calE_f)=2S(\rho_\calS)$. If neither condition is satisfied, then $I(\calS:\calE_f)=0$.
The probabilities associated with the two conditions are independent from one another, and can thus be computed separately.

We will think of the environment as being made of boxes, each box can contain at most $k$ qubits, and the total number of boxes is $N$. We randomly fill the boxes with $m$ qubits, where $m$ is the size of the environmental fraction. Condition A corresponds to the case where at least one box is completely full. Condition B correspond to the case where no box is empty. The issue now is understanding the probabilities of the boxes being empty or full when we randomly fill them with qubits.

We wish to know how many configurations correspond to having at least one box full, $\mathcal{N}_A$.
We assume that we entirely fill one box beforehand, and distribute the remaining qubits in the remaining free slots.
The number of possible combinations is $\binom{Nk-k}{m-k}$, but we must also take into account that the initial choice of the full box was arbitrary, so that the total number of combinations is $N\binom{Nk-k}{m-k}$. However, this corresponds to an over-counting, because we are counting multiple times the cases where more than one box is completely filled. We correct this over-counting by subtracting all the possible combinations where two boxes are completely filled beforehand, keeping in mind that this also corresponds to an over-counting of the number of events with more than two boxes completely filled, and so on.
The total number of configurations that fulfill condition A results in a sum of alternating sign, that reads
\begin{equation}
    \mathcal{N}_A=\sum_{j=1}^{N}(-1)^{1+j}\binom{N}{j}\binom{Nk-jk}{m-jk}.
\end{equation}
The total number of possible configurations is simply $\mathcal{N}=\binom{Nk}{m}$, so that the probability that condition A is fulfilled is $p_A=\frac{\mathcal{N}_A}{\mathcal{N}}$.

Instead of computing the number of configurations where none of the boxes are empty, $\mathcal{N}_B$, it's easier to compute the number of configurations where there is at least an empty box, $\mathcal{N}_{\overline{B}}$, by following a similar reasoning to the previous case: we keep an entire box empty, and then distribute the qubits in the remaining boxes, also in this case we need to take care of the resulting over-counting, leading to 
\begin{equation}
    \mathcal{N}_{\overline{B}}=\sum_{j=1}^{N}(-1)^{1+j}\binom{N}{j}\binom{Nk-jk}{m}.
\end{equation}
Since we are interested in the number of configurations where none of the boxes are empty, we have that $\mathcal{N}_B=\mathcal{N}-\mathcal{N}_{\overline{B}}$, and the probability that condition B is fulfilled is $p_B=\frac{\mathcal{N}_B}{\mathcal{N}}$.

Based on these results we can finally compute the averaged mutual information between the system $\calS$ and an environmental fraction $\calE_f$ as $\Tilde{I}(\calS:\calE_f)=(p_A+p_B)S(\rho_\calS)$.

If we want to compute the averaged accessible information $I_{\operatorname{acc}}(\calS:\calE_f)$, we should take into account that when both conditions A and B are satisfied, the accessible information is only $I_{\operatorname{acc}}(\calS:\calE_f)=S(\rho_\calS)$. therefore
\begin{equation}
    \Tilde{I}_{\operatorname{acc}}(\calS:\calE_f)=(p_A+p_B-p_Ap_B)S(\rho_\calS)
\end{equation}
\section{Mutual information for W encodings}

Here we consider all possible states resulting from tracing out environmental components from the global state, when information about the system is encoded in W-like states, and compute the QMI between the system and the resulting environmental fraction. In the case of W encodings, it is not trivial to compute the accessible information, we will therefore compute the QMI, using it as an upper bound for the accessible information.

For simplicity, we will focus on the case of $k=3$. The global state (\cref{eq:W_encoding} in the main text) is
\begin{equation}
    \ket{\Psi}=\sqrt{p}\ket{0}\otimes\ket{w^{+}}^{\otimes N}+\sqrt{1-p}\ket{1}\otimes\ket{w^{-}}^{\otimes N}.
    \label{eq:global}
\end{equation}
Where $\ket{w^+}=\frac{1}{\sqrt{3}}(\ket{001}+\ket{010}+\ket{100})$ and $\ket{w^-}=\frac{1}{\sqrt{3}}(\ket{001}+\omega_3\ket{010}+\omega_3^2\ket{100})$ are the states in~\cref{eq:W_states}, with $\omega=e^{2\pi i/3}$.
We will use a compact notation with $\ket{\Tilde{0}}_N=\ket{0}\otimes\ket{w^{+}}^{\otimes N}$ and $\ket{\Tilde{1}}_N=\ket{1}\otimes\ket{w^{-}}^{\otimes N}$.

As with the GHZ-like encoding, depending on which qubits are traced out, we will compute the resulting QMI for four different cases.

\subsection{First case}
We will first consider the case where at least one W state has been fully traced out. In this case, if two qubits are traced out from a W state, the last one will be in a product state with everything else, and will not influence the QMI because it will cause the same increase of entropy for both $\rho_{\mathcal{S}\mathcal{E}_f}$ and $\rho_{\mathcal{E}_f}$. When we trace out a qubit from a W state, we create a mixture between two pure states. More specifically, if we trace out the $i$-th qubit $\mathrm{tr}_{i}\{\ketbra{w^{\pm}}\}=\rho^{(i)}_{\pm}=1/k\ketbra{0}^{\otimes k-1}+\frac{k-1}{k}\ketbra{w^{\pm}_{i}}$, with $\ketbra{0}^{\otimes k-1}$ and $\ketbra{w^{\pm}_{i}}$ orthogonal to one another. 

When we trace out one W state from the global state, but at least one W state remains intact, the states are the following:
\begin{equation}
    \rho_{\mathcal{S}\mathcal{E}_f^{(0)}}=p\ketbra{\Tilde{0}}_{n}\otimes{\rho_{+}^{(i)}}^{\otimes m}+(1-p)\ketbra{\Tilde{1}}_n\otimes{\rho_{-}^{(i)}}^{\otimes m},
\end{equation}
\begin{equation}
    \rho_{\mathcal{E}_f^{(0)}}=p\ketbra{w^{+}}^{\otimes n}\otimes{\rho_{+}^{(i)}}^{\otimes m}+(1-p)\ketbra{w^{-}}^{\otimes n}\otimes{\rho_{-}^{(i)}}^{\otimes m},
\end{equation}
It is actually not needed to compute the entropies of the states above. The two states clearly have the same entropy, so the mutual information is $I(\calS:\calE_f^{(0)})=S(\rho_\calS)$.

\subsection{Second case}
If at least one group is traced out entirely, and all other groups have been traced out partially, the states are the following:
\begin{equation}
    \rho_{\mathcal{S}\mathcal{E}_f^{(1)}}=p\ketbra{0}\otimes{\rho_{+}^{(i)}}^{\otimes m}+(1-p)\ketbra{1}\otimes{\rho_{-}^{(i)}}^{\otimes m},
    \label{eq:2nd_state}
\end{equation}
\begin{equation}
    \rho_{\mathcal{E}_f^{(1)}}=p{\rho_{+}^{(i)}}^{\otimes m}+(1-p){\rho_{-}^{(i)}}^{\otimes m}
    \label{eq:2nd_environment},
\end{equation}

We can explicitly rewrite~\cref{eq:2nd_state} in the following way
    \begin{equation}
    \small
    \rho_{\mathcal{S}\mathcal{E}_f^{(1)}}=
        p\ketbra{0}\otimes\Bigg[1/k\ketbra{0}^{\otimes k-1}+(k-1)/k\ketbra{w^{+}_{i}}\Bigg]^{\otimes m}+(1-p)\ketbra{1}\otimes\Bigg[1/k\ketbra{0}^{\otimes k-1}+(k-1)/k\ketbra{w^{-}_{i}}\Bigg]^{\otimes m},
    \end{equation}
    all the terms in the above states are pure and orthogonal to one another, so the resulting entropy is only due to classical mixing. The entropy of this state results from the probability vector $(p, 1-p)\otimes(1/k, (k-1)/k)^{\otimes m}$, whose entropy is straightforward to compute.

In the case of~\cref{eq:2nd_environment} we end up having states of the form $p\ketbra{w^{+}_{i}}^{\otimes m}+(1-p)\ketbra{w^{-}_{i}}^{\otimes m}$, the eigenvalues of such states are $\lambda=(1\pm\sqrt{1+4p(1-p)(d-1)})/2$ with $d=|\bra{w^{+}_i}^{\otimes m}\ket{w^{-}_i}^{\otimes m}|^{2}$, in the case of $p=1/2$ they reduce to $\lambda=(1\pm \sqrt{d})/2$.
    
    The entropy for the state in~\cref{eq:2nd_environment} is thus
    \begin{equation}
        S[(1/3, 2/3)^{\otimes m}]+\sum_{j=0}^{m}(1/3)^{m-j}(2/3)^{j}\binom{m}{j}S(p\ketbra{w^+_i}^{\otimes j}+(1-p)\ketbra{w^-_i}^{\otimes j}),
    \end{equation}
    where $S[(1/3, 2/3)^{\otimes n}]$ is the entropy of the probability vector $(1/3, 2/3)^{\otimes n}$. The QMI is thus
    \begin{equation}
        I(\calS:\calE_f^{(1)})=H'(m)=\sum_{j=0}^{m}(1/3)^{m-j}(2/3)^{j}\binom{m}{j}S(p\ketbra{w^+_i}^{\otimes j}+(1-p)\ketbra{w^-_i}^{\otimes j}).
    \end{equation}
    Notice how it is necessary to know the number of W states from which only one qubit has been traced out.
    
    \subsection{Third case}
    Let's now focus on the alternate scenario, where we never entirely trace out a W-group.
    We start once again from the global state~\cref{eq:global}, assuming $p=\frac{1}{2}$
    \begin{equation}
        \frac{1}{2}[\ket{\Tilde{0}}_{N-1}\ket{w^{+}}+\ket{\Tilde{1}}_{N-1}\ket{w^{-}}][\bra{\Tilde{0}}_{N-1}\bra{w^{+}}+\bra{\Tilde{1}}_{N-1}\bra{w^{-}}].
    \end{equation}
    
    After tracing out one qubit we obtain the following state:
    \begin{equation}
        \frac{1}{3}[\ket{\Tilde{0}}_{N-1}+\ket{\Tilde{1}}_{N-1}][\bra{\Tilde{0}}_{N-1}+\bra{\Tilde{1}}_{N-1}]\ketbra{0}^2+\frac{2}{3}[\ket{\Tilde{0}}_{N-1}\ket{w_{i}^{+}}+\ket{\Tilde{1}}_{N-1}\ket{w_{i}^{-}}][\bra{\Tilde{0}}_{N-1}\bra{w_{i}^{+}}+\bra{\Tilde{1}}_{N-1}\bra{w_{i}^{-}}],
    \end{equation}
    so we have a statistical mix of pure distinguishable states. So long as we trace out at most one qubit from each W-group, the procedure can be easily generalised and we end up with a sum of pure distinguishable states where $(1/3,\ 2/3)^{\otimes m}$ is the resulting probability vector from which we can compute the entropies.
    To be precise, relative phases may come up depending on which qubit is traced out. As we keep tracing out qubits, these phases may build up, making the exact writing of the states non trivial. However, these relative phases are of no consequence for us as we are always left with distinguishable states (notice that in the previous cases such phases did not exist, since we always had at least one fully traced-out W state). 
    This is no longer true when we trace out more than one qubit from each W-group, as in this case such relative phases influence the entropy of the resulting states.
    
    To illustrate this, we start from the initial state
    \begin{equation}
        \ket{\Psi}=\ket{\Tilde{0}}_{N}+\ket{\Tilde{1}}_{N},
    \end{equation}
    after tracing out one qubit we have (from now on, we will use $\mathbb{P}\left(\ket{\psi}\right)$ as a compact notation for $\ketbra{\psi}$)
    \begin{equation}
        \frac{1}{3}\mathbb{P}\left(\ket{\Tilde{0}}_{fN}\ket{0}+\phi_{i}^{0}\ket{\Tilde{1}}_{fN}\ket{0}\right)+
        \frac{2}{3}\mathbb{P}\left(\ket{\Tilde{0}}_{fN}\ket{w^{+}_{i}}+\ket{\Tilde{1}}_{fN}\ket{w^{-}_{i}}\right).
    \end{equation}
    $\phi_{i}^{0}$ is the relative phase of the $i$-th qubit of the $\ket{w^{-}}$ state, the superscript $0$ indicates that it refers to the first W state we are tracing out. To be precise, $[\phi_{0}^{j},\ \phi_{1}^{j},\ \phi_{2}^{j}]=[1,\ e^{i\frac{2}{3}\pi},\ e^{-i\frac{2}{3}\pi}]$. 
    If we trace out another qubit from another W state we have
    \begin{equation}
    \begin{split}
        &\frac{1}{9}\mathbb{P}\left(\ket{\Tilde{0}}_{fN}\ket{00}+\phi_{i}^{0}\phi_{j}^{1}\ket{\Tilde{1}}_{fN}\ket{00}\right)+
        \frac{2}{9}\mathbb{P}\left(\ket{\Tilde{0}}_{fN}\ket{0w^{+}_{j}}+\phi_{i}^{0}\ket{\Tilde{1}}_{fN}\ket{0w^{-}_{j}}\right)+\\
        &\frac{2}{9}\mathbb{P}\left(\ket{\Tilde{0}}_{fN}\ket{w^{+}_{i}0}+\phi_{j}^{1}\ket{\Tilde{1}}_{fN}\ket{w^{-}_{i}0}\right)+
        \frac{4}{9}\mathbb{P}\left(\ket{\Tilde{0}}_{fN}\ket{w^{+}_{i}w^{+}_{j}}+\ket{\Tilde{1}}_{fN}\ket{w^{-}_{i}w^{-}_{j}}\right),
    \end{split}
    \end{equation}
    See how this procedure always keeps the resulting states distinguishable, so that the entropy of the resulting state is only due to the relative probabilities.
    Now we trace another qubit from one of the surviving pairs, resulting in
    \begin{equation}
    \footnotesize
    \begin{split}
        &\frac{1}{9}\mathbb{P}\left(\ket{\Tilde{0}}_{fN}\ket{00}+\phi_{i}^{0}\phi_{j}^{1}\ket{\Tilde{1}}_{fN}\ket{00}\right)+
        \frac{1}{9}\mathbb{P}\left(\ket{\Tilde{0}}_{fN}\ket{00}+\phi_{i}^{0}\phi_{j'}^{1}\ket{\Tilde{1}}_{fN}\ket{00}\right)+
        \frac{1}{9}\mathbb{P}\left(\ket{\Tilde{0}}_{fN}\ket{01}+\phi_{j}^{1}\phi_{j''}^{1}\ket{\Tilde{1}}_{fN}\ket{01}\right)\\
        +&\frac{2}{9}\mathbb{P}\left(\ket{\Tilde{0}}_{fN}\ket{w^{+}_{i}0}+\phi_{j}^{1}\ket{\Tilde{1}}_{fN}\ket{w^{-}_{i}0}\right)+
        \frac{2}{9}\mathbb{P}\left(\ket{\Tilde{0}}_{fN}\ket{w^{+}_{i}0}+\phi_{j'}^{1}\ket{\Tilde{1}}_{fN}\ket{w^{-}_{i}0}\right)+
        \frac{2}{9}\mathbb{P}\left(\ket{\Tilde{0}}_{fN}\ket{w^{+}_{i}1}+\phi_{j''}^{1}\ket{\Tilde{1}}_{fN}\ket{w^{-}_{i}1}\right).
    \end{split}
    \end{equation}
    In this case some of the resulting projectors are not distinguishable from one another anymore, resulting in mixed states. The coherence factor of the resulting mixed projector will be $\frac{1}{2}(\phi_{j}^{1}+\phi_{j'}^{1})$, whose absolute value is $\frac{1}{2}$ $\forall j\neq j'$. By tracing out a qubit from the other W state we have
    
    \begin{equation}\footnotesize
    \begin{split}
        &\frac{1}{9}\mathbb{P}\left(\ket{\Tilde{0}}_{fN}\ket{00}+\phi_{i}^{0}\phi_{j}^{1}\ket{\Tilde{1}}_{fN}\ket{00}\right)+
        \frac{1}{9}\mathbb{P}\left(\ket{\Tilde{0}}_{fN}\ket{00}+\phi_{i}^{0}\phi_{j'}^{1}\ket{\Tilde{1}}_{fN}\ket{00}\right)+
        \frac{1}{9}\mathbb{P}\left(\ket{\Tilde{0}}_{fN}\ket{01}+\phi_{j}^{1}\phi_{j''}^{1}\ket{\Tilde{1}}_{fN}\ket{01}\right)\\
        +&\frac{1}{9}\mathbb{P}\left(\ket{\Tilde{0}}_{fN}\ket{00}+\phi_{i'}^{0}\phi_{j}^{1}\ket{\Tilde{1}}_{fN}\ket{00}\right)+
        \frac{1}{9}\mathbb{P}\left(\ket{\Tilde{0}}_{fN}\ket{00}+\phi_{i'}^{0}\phi_{j'}^{1}\ket{\Tilde{1}}_{fN}\ket{00}\right)+
        \frac{1}{9}\mathbb{P}\left(\ket{\Tilde{0}}_{fN}\ket{01}+\phi_{i'}^{0}\phi_{j''}^{1}\ket{\Tilde{1}}_{fN}\ket{01}\right)\\
        +&\frac{1}{9}\mathbb{P}\left(\ket{\Tilde{0}}_{fN}\ket{10}+\phi_{i''}^{0}\phi_{j}^{1}\ket{\Tilde{1}}_{fN}\ket{10}\right)+
        \frac{1}{9}\mathbb{P}\left(\ket{\Tilde{0}}_{fN}\ket{10}+\phi_{i''}^{0}\phi_{j'}^{1}\ket{\Tilde{1}}_{fN}\ket{10}\right)+
        \frac{1}{9}\mathbb{P}\left(\ket{\Tilde{0}}_{fN}\ket{11}+\phi_{i''}^{0}\phi_{j''}^{1}\ket{\Tilde{1}}_{fN}\ket{11}\right).
    \end{split}
    \end{equation}
    Once again some projectors are not distinguishable from one another, resulting in mixed states. The coherence factor of the resulting projectors is either $\frac{1}{2}(\phi_{j}^{1}+\phi_{j'}^{1})$ or $\frac{1}{4}(\phi_{i}^{0}+\phi_{i'}^{0})(\phi_{j}^{1}+\phi_{j'}^{1})$. In general the absolute value of the resulting coherence factor will be $\frac{1}{2^{l}}$, where $l$ is the number of $\ket{0}$ states corresponding to W states were two qubits have been traced out. Interestingly enough, the entropy of such states is the same one of the states of the form $\frac{1}{2}\mathbb{P}\left(\ket{w^+_i}^{\otimes l}\right)+\frac{1}{2}\mathbb{P}\left(\ket{w^-_i}^{\otimes l}\right)$.
    
    To sum things up, whenever we have a state where we have not traced out more than one qubit per pair, then the probability vector of the resulting state is $(1/3, 2/3)^{\otimes m}$. All the states are pure and distinguishable, so the entropy is only the entropy resulting from the probability. If we further trace out qubits from the surviving W states, we end up with a probability vector of the form $(2/3, 1/3)^{\otimes k}\otimes (1/3, 2/3)^{\otimes m-k}$, where $k$ is the number of W states from which we have traced out two qubits, and $m-k$ is the number of W states from which we have traced out only one qubit. So actually, while the individual probabilities have changed, the entropy of the probability vector remains the same. However, now the states are still distinguishable but not necessarily pure. The mixed states are mixtures with a coherence factor of $\frac{1}{2^{l}}$ with $0\leq l\leq k$.
    The entropy of the state $\rho_{\calS\calE_f^{(2)}}$ is thus
    \begin{equation}
    \footnotesize
        S(\rho_{\calS\calE_f^{(3)}})=S[(1/3, 2/3)^{\otimes n}]+\sum_{i=0}^{m-k}(1/3)^{i}(2/3)^{m-k-i}\binom{m-k}{i}\sum_{j=0}^{k}(2/3)^{k}(1/3)^{k-j}\binom{k}{j}S\left(\frac{1}{2}\mathbb{P}\left(\ket{w^+_i}^{\otimes j}\right)+\frac{1}{2}\mathbb{P}\left(\ket{w^-_i}^{\otimes j}\right)\right).
        \label{eq:entropy_third}
    \end{equation}
    We will refer to the second term of~\cref{eq:entropy_third} as $\Tilde{H}$.
    When we trace out the system as well, all the relative phases are lost, and we end up once again with a collection of pure distinguishable states with a resulting probability vector of $(p, 1-p)\otimes(1/3,\ 2/3)^{\otimes n}$, this leads to a mutual information of $I(\calS:\calE_f^{(2)})=2S(\rho_{\mathcal{S}})-\Tilde{H}$.
    
    \subsection{Fourth case}
    In the fourth and final case, at least one ancilla has been traced out from all groups, but no group has been entirely traced out.
    The state structure of the resulting state is exactly the same as in the previous scenario, so that the entropy is also the same.
    Instead, when tracing out the system as well the state structure is the same one as~\cref{eq:2nd_environment}.
    Combining our previous results, the mutual information in this case is $I(\calS:\calE_f^{(3)})=S(\rho_\calS)+ H'-\Tilde{H}$, where we need to keep track of the number of surviving pairs as well as the number of surviving singlets.
    
\section{Combinatorial computation for W encodings}
In the case of the W states, the combinatorial calculation is now more complicated compared to the GHZ encoding. Using the parallelism introduced earlier, it is no longer sufficient to know the probabilities for the boxes to be empty or full, but we also need to know the probabilities for a certain number of boxes to have exactly one qubit or two qubits, as this now influences the corresponding mutual information. 

We can approach the combinatorial calculation in this way: once we fix the number of qubits of the environmental fraction $m$, we can go through all the possible integer partitions of that specific number (for example, two integer partitions of $5$ are $1^{1}2^{2}$ and $1^{5}$), these partitions correspond to the possible ways in which we can distribute the qubits in different boxes. Some partitions however are not compatible with the physical problem at hand: for example, we cannot have integers larger than $k$, and the total number of integers making up our partitions cannot be larger than $N$. We may also wish to impose extra conditions: for example, if we wish that no W state is fully selected, we can impose that no integers in the partitions should be larger than $k-1$, and if we wish that no W state is entirely traced out, we can impose that the total number of integers in the partition is exactly $N$. If a partition $\mathbf{i}$ is $1^{i_{1}}2^{i_{2}}...$ the number of possible combinations is given by the polynomial $\binom{\sum_{j} i_{j}}{i_{0},i_{1},i_{2}\dots}$, if we also consider the degeneracy resulting from shuffling the qubits inside a box, we have that for every integer partition $\mathbf{i}$ of $m$, the number of possible configurations is 
\begin{equation}
    \mathcal{N}_{\mathbf{i}}=\binom{\sum_{j} i_{j}}{i_{0},i_{1},i_{2}\dots}\Pi_{j}\binom{k}{j}^{i_{j}},
\end{equation}
with an associated probability $p_{\mathbf{i}}=\frac{\mathcal{N_{\mathbf{i}}}}{\mathcal{N}}$

\end{document}